# Performance Analysis of Uplink & Downlink Transmission in CDMA System

Md. M. Hossain, Md. M. Rahman, and Md. A. Alim

**Abstract**—CDMA is a multiple access method in which the user's uses spread spectrum techniques and occupy the entire spectrum whenever they transmit. In wireless communication signal-to-noise ratio (SNR) is the very important parameter that influences the system performance. Any mode of mobile transmission is not free from channel impairment such as noise, interference and fading. This channel impairment caused signal distortion and degradation in SNR.Also there are differences between uplink (forward channel) and downlink (reverse channel).Along with these differences, both the links use different codes for chanellizing the individual users. This paper simulates the expressions for the pdfs of the SNR for both uplink and downlink transmission assuming that the system is operating at an average signal-to-noise ratio is 6dB per information bit,

———————————— ◆ ————————————

## 1 INTRODUCTION

IN CDMA system there is a difference between uplink (forward channel) and downlink (reverse channel), ranging from modulating scheme to error correcting codes. The purposes of error correcting codes when applied to channel coding, improve the error performances of the system. There are two major error correcting codes used in channel coding block codes and convolution codes. Block codes, as name implies an information sequence one block at a time. Convolution codes have a memory property which depend on constraint length k which is very good impact on both uplink and downlink transmission. Along with this difference, both the links use different codes for canalizing the individual users.

## 2 SYSTEM MODEL

A block scheme of the transmitter is shown in Fig-1(a). The information bit stream of the $k^{th}$ user, $b_k (1) \in \{0,1\}$, convolutionally encoded by the encoder CC and interleaved by the interleaver $\prod$. The resulting sequence is QPSK-modulated and finally scrambled with $q_k (n)$. The purpose of this scrambling is to make the interfering user signals look random in the receiver. The receiver has the error correcting ability of the convolutional code. In Fig. 1(b) a flat fading channel is introduced. The receiver does not rely on orthogonally to distinguish between different user signals, but solely on the error correcting ability of the convolutional code. The channel coefficient $a[n] = a[n]ei <![n]$, is modeled as a circular complex Gaussian random process, i.e., $a[n]$ is Rayleigh distributed and uniformly distributed on $\{O,2J1\}$. The time-varying channel tap is normalized so that the average power after the channel equals the transmitted power $E[!a[n]r] = 1$. The received signal is descrambled and phase-aligned, demodulated, dein-terleaved and finally decoded to yield an estimate of the transmitted bit sequence $b \sim_k$ [1]. QPSK-demodulator produces a sequence of real numbers by extracting the real and imaginary parts of the sequence input to the block. Since the data stream is convolutionally encoded the optimum sequence detector in the single-user case is the maximum likelihood (ML) sequence detector which is

Fig. 1. Discrete-time CDMA System Model

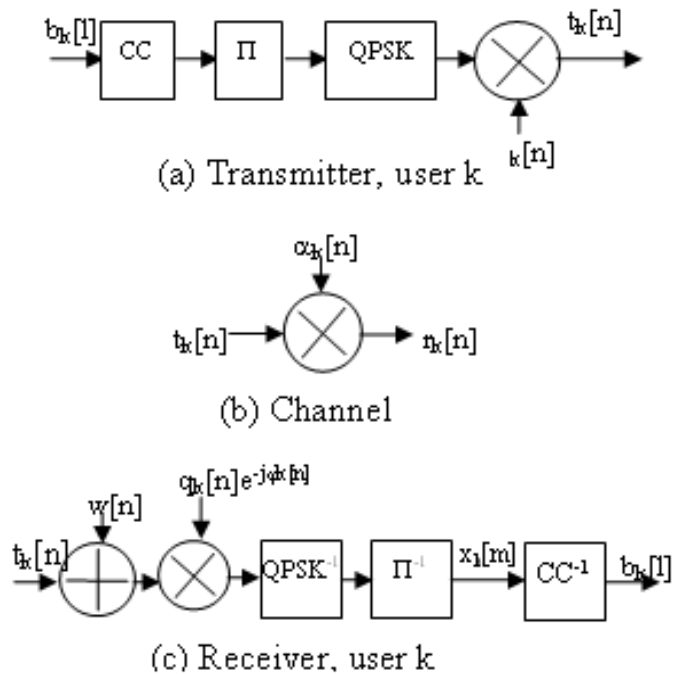

(a) Transmitter, user k

(b) Channel

(c) Receiver, user k

about choosing the estimate of transmitted code vector for which the log-likelihood function is maximized, at the decod-

————————————————
• Md. M. Hossain is with the Electronics and Communication Engineering Discipline, Khulna University, Khulna - 9208, Bangladesh.
• Md. M. Rahman is with the Electronics and Communication Engineering Discipline, Khulna University, Khulna - 9208, Bangladesh.
• Md. A. Alim is with the Electronics and Communication Engineering Discipline, Khulna University, Khulna - 9208, Bangladesh.

ing end. The receiver structure is shown in Fig. 1(c).

## 3 ANALYSIS

In a downlink scenario the desired signal as well as the intra-cell interference pass through the same channel and arrive at the receiver in phase (within one ray). However in an uplink scenario different user signals, $t_k[n]$, pass through independently fading channels and arrive with independent phases. The interference level in an uplink system using BPSK-modulation is therefore reduced compared to a downlink system. Therefore, to make a fair comparison between the up- and downlink systems QPSK-modulation is assumed. A reasonable assumption in the analysis of an uplink CDMA system is perfect power control in the average sense, i.e., all user signals arrive at the base station having the same average power. In a downlink system this is almost certainly not the case. Since different mobile stations are located at different positions, they will request different transmitted powers in order to maintain a given quality of service. However, since the aim of the analysis is to explain the performance deference between CDMA in the uplink compared to the downlink, it is illustrative to assume that both the desired signal and each interfering user have the same average power also in the downlink. For the same reason synchronous transmission is assumed both in the up and downlink. The scrambling sequences, $q_k[n]$, are sequences of independent, identically distributed random variables taking on the values $\pm 1$ with probability $1/2$. For both the uplink and the downlink analysis but in computer simulations perfect channel estimates, interleaving and synchronization are assumed. When evaluating the performance of a convolutionally coded code it is customary to upper bound the BER by the union bound [4]-[2]

$$p_b \leq \sum_{d=d_f}^{\infty} c_d \; p_2(d)$$  (1)

where $d_f$ is the free distance of the code, $p_2(d)$ is the pair-wise error probability of two sequences separated by hamming distance $d$ and $c_d$ are the information error weights of the code. An approximate value of the BER may be obtained by truncating the sum in equation (1) at some suitable value of d. To find an expression for the error probability in a fading scenario, the standard approach is to obtain an expression for the error probability given the instantaneous signal-to-noise ratio and then average this with respect to the probability density function. Therefore, writing $p_r(e)$ for the probability of error, then we have

$$p_r(e) = \int_{-\infty}^{\infty} f(\gamma) \, p_r(e/\gamma) d\gamma$$  (2)

where the symbols have their usual meanings and if the interference is additive Gaussian noise, the probability of error given the instantaneous signal-to-noise ratio expressed in terms of the Q-function is

$$p_r(e \mid \gamma) = Q(\sqrt{2\gamma})$$  (3)

$$Q(x) = \frac{1}{\sqrt{2\pi}} \int_{-\infty}^{\infty} e^{-2}/2 \, dt$$  (4)

therefore, to evaluate $p_2(d)$ the expression for the pdf of the signal-to-noise ratio of d combined codes bits must be obtained

for both uplink and downlink

### 3.1 Uplink and Downlink

Over the uplink different user signals $t_k[n]$, pass through independent channels. The equivalent average signal-to-noise ratio $\overline{\gamma_c}$ ,of the coded bits is given by [4]

$$\overline{\gamma_c} = \frac{\overline{\xi}}{2(k-1)+N_o}$$  (5)

while $\xi$ denote received average energy per coded bit, No denote noise, k denote constraint length. Since the equivalent instantaneous signal-to-noise ratio, $\gamma_c[m]$ of each coded bits is $X^2$–distributed with 2 degrees of freedom,

$$f_{\gamma_c}(\gamma_c) = e^{-\gamma_c/\gamma_c}/\gamma_c^-$$  (6)

The equivalent instantaneous signal-to-noise ratio $\gamma[m]$ of d combined coded bits will also be $X^2$-distributed, but with 2d degrees of freedom [5],

$$f_\gamma(\gamma) = \gamma^{d-1}e^{-\lambda}_{\gamma_c}/(d-1)!\,\gamma_c^{-d}$$  (7)

Now for the downlink the equivalent signal-to-noise ratio is independent from sample to sample and the pdf of the equivalent signal-to-noise ratio of d combined samples thus equal

$$f_\tau(\gamma) = f_{\tau_c}(\gamma) * f_{\tau_c}(\gamma) * \cdots\cdots * f_{\gamma_c}(\gamma)$$  (8)

The free distance $d_f$ of the code must be large to yield acceptable performance at the low signal-to-noise ratios that result from the multi-user interference. Typically, $d_f$ is more than 100 for the codes of interest. Thus, to evaluate the union bound the pair-wise error probabilities, $p_2(d)$, are only required for d > 100. If d is this large the central limit theorem may be applied to reasonably approximate the pdf of signal-to-noise ratio by a Gaussian pdf [3].

$$f\tau(\gamma) = Ae^{-\frac{(\gamma-\mu)}{2\sigma^2}}/\sqrt{2\pi}\sigma \, , \quad 0 < \gamma < \frac{d}{2(k-1)}$$

$$f\tau(\gamma) = 0 \text{ , elsewhere}$$  (9)

### 3.2 RESULTS

This paper worked out a MathCAD simulating program for the CDMA system using QPSK-modulating and a rate 1/4, constraint length 10 ,maximum free distance convolutional code for spreading [7].The free distance, $d_f$, of this code is 164. Assuming that the system is operating at an average signal-to-thermal noise ratio $\xi / N_o$ of 6 dB per information bit. In fig 2,3,4 the pdf of the signal-to-noise ratio per coded bit and the pdf of $d_f = 164$ combined coded bits are shown for both up- and downlink by using equation 6 and 9.

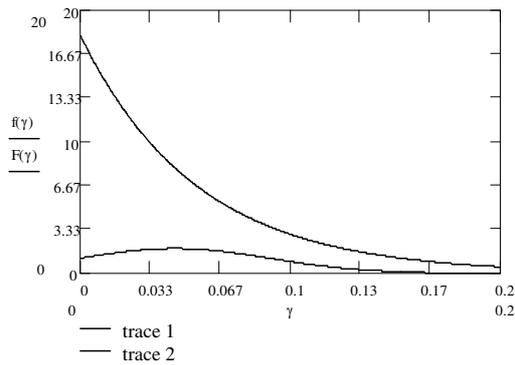

Fig. 2. pdf of signal-to-noise ratio for both uplink and downlink when k=10

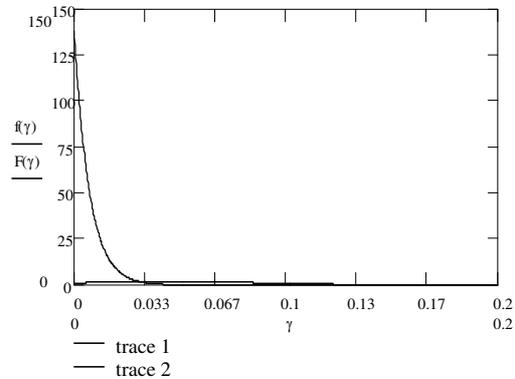

Fig. 3. pdf of signal-to-noise ratio for both uplink and downlink when k=70

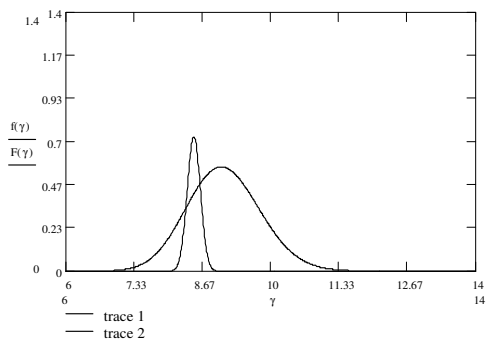

Fig. 4. pdf of signal-to-noise ratio for both uplink and downlink when k=10 and $d_f$ =164

It is clear from Fig. 2-4 that the pdfs of signal-to-noise ratio differs substantially between uplink and downlink.

## 4 CONCLUSION

We dealt with the error performance criterion on the basis of modulation techniques used in both the links uplink and down link scenario. Expression simulate in both cases with respect to signal-to-noise ratio. Here one could obtain good codes and characteristics of a good code. Free distance and distance spectrum are the parameters that determine the error performance of a code. In turn these two parameters are dependent on code rate and length. So it was concluded that the codes with lower code rates and length were the best approximation. However it was further illustrated that after attaining certain limits, lowering the rate was equivalent to repeating the parent codes. Hence use of a code of rather high rate combined with a simple repetition encoder can perform almost equally well as a code of much lower rate. This observation is important for designing efficient synchronization algorithms and also for efficient decoder implementation.

## REFERENCES

[1] A. Persson, J. Lassing, T. Ottosson, and E. Str.'om, "On the differences between uplink and downlink transmission in code-spread CDMA systems," in Proc. *IEEE Vehicular Technology Conference Spring, Rhodes, Greece, May 2001,* vol. 4, pp. 2421- 2425.

[2] G. E. P. Box, W. G. Hunter, and J. S. Hunter, Statistics for Experimenters: *An Introduction to Design, Data Analysis and Model Building,* John Wiley & Sons, 1978.

[3] G. R. Grimmett and D. R. Stirzaker, *Probability and Random Processes,* Oxford Science Publications, 2nd edition, 1991.

[4] J. G. Proakis, *Digital Communications,* 3rd edition, McGraw-Hill, 1995. [9] A. Viterbi and J. K. Omura, *Principles of digital communication and coding,*McGraw-Hill, New-York, 1979.

[5] J. Lassing, T. Ottosson, and E. Strom, "On the union bound applied to convolutional codes," in Proc. *IEEE Vehicular Technology Conference Fall, Atlantic City, USA, Oct 2001.*

[6] P. D. Papadimitriou and C. N. Georghiades, "On asymptotically optimum rate 1/n convolutional codes for a given constraint length," *IEEE Communications Letters,* vol. 5, no. 1, pp. 25-27, Jan. 2001.

[7] P. Frenger, P. Orten, and T. Ottosson, "Convolutional codes with optimum distance spectrum," *IEEE Communications Letters,* vol. 3, no. 11, pp. 31.7-319, Nov. 1999.

**Md. M. Hossain** received his B.Sc Engineering degree in Electronics and Communication in the year of 2003 from Khulna University, Khulna-9208, Bangladesh. He is now the faculty member of Electronics and Communication Enfgineering Discipline, Khulna University, Khulna-9208, Bangladesh since 2005. His current research interest is wireless communication, modulation, channel coding and fading. His numbers of published papers are 3 among them one was published in the proceding of 12[th] ICCIT, 2009, Dhaka, Bangladesh. Another one was published in the proceding of 2[nd] international conference on ICEESD, 2009, DHAKA, Bangladesh and last one was published in the conference on Engineering Research, Innovation and Education, 2010, SUST, Sylhet, Bangladesh.

**Md. M. Rahman** received his B.Sc. Engineering degree in Electronics and Communication Engineering with the highest CGPA in his batch in the year of 2003 from Khulna University, Khulna-9208, Bangladesh. He is working as a faculty member of Electronics and Communication Engineering Discipline, Khulna University, Khulna-9208, Bangladesh since 2003. His current research interest is in Telecommunication, Opto-Electronic Devices and Communication, OFDMA.

**Md. A. Alim** Assistant Professor in Electronics and Communication Engineering in Khulna University, Bangladesh. His research interest includes wireless communication, mobile communication and channel coding. Before joining in Khulna University, he has professional experience and training from SIEMENS Bangladesh as an executive engineer in the arena of mobile communication.